
\input harvmac

%

\setbox\strutbox=\hbox{\vrule height12pt depth5pt width0pt}

\def\strut{\relax\ifmmode\copy\strutbox\else\unhcopy\strutbox\fi}

\nref\raug{St. Augustine, {\it DeGenesi ad Litteram}, Book II, xviii, 37.}
\nref\reras{D. Erasmus,{\it In Praise of Folly}, (Princeton
Univ. Press, 1941), Trans. H.H. Hudson, pp.76-77.}
\nref\rptol{Ptolemy, {\it The Almagest} (Springer--Verlag 1984),
Trans. G.J. Toomer.}
\nref\rdesc{R. Descartes, {\it Discourse on Method}, in {\it Descartes,
Philosophical Writings}, Modern Library
(Random House 1956) p. 106.}
\nref\rbaca{F. Bacon, {\it Novum Organum}, Open Court (Chicago 1994)
ed. P. Urbach and J. Gibson, pp. 53-68.}    
\nref\rari{Aristotle, {\it Metaphysics}, $1026^a$ 19-20, in {\it A New
Aristotle Reader}, (Princeton Univ. Press,1987) ed. J.L. Ackrill, p. 280.}
\nref\raqu{T. Aquinas, {\it Expositio supra Lebrum Boethii de
Trinitate}, in {\it Thomas Aquinas Selected Philosophical Writings}
(Oxford Univ. Press, 1993) pp. 1-50}
\nref\rgala{Galileo Galilei, {\it Siderius  Nuncius}, Univ. Chicago Press,
1989), ed. Albert van Helden, pp.35-30.}
\nref\rbacb{{F. Bacon,\it op. cit.} pp. 143-144.}
\nref\rbacd{{F. Bacon,\it op. cit.} pp. 142.}
\nref\rbace{{F. Bacon,\it op. cit.} pp. 143.}
\nref\rvol{Voltaire, {\it Letters on England}, letter 15 {\it On the
System of Gravitation},( Dover, 1980) pp. 73-- 81.}
\nref\rleib{G.W. Leibnitz,{\it Reply to the fourth letter of Clarke},
in {\it Leibnitz Philosophic Works} ed. G.H. Parkinson, Everyman
(1973) pp. 221-238.}
\nref\rrog{A.K. Rogers, {\it A Students History of Philosophy},
(MacMillan, 1901) p.436}
\nref\rym{C.N. Yang and R.L. Mills, {\it Conservation of Isotopic Spin
and Isotopic Gauge Invariance}, Phys. Rev. 96 (1954) 191.}
\nref\rly{T.D. Lee and C.N. Yang, {\it Question of Parity conservation
in Weak Interactions}, Phys. Rev. 104 (1956) 254.}
\nref\rschwa{J. Schwinger, {\it Selected Papers in Quantum
Electrodynamics}, (Dover 1958) p. xvii}
\nref\rschwb{J. Schwinger,{\it Gauge Invariance and Mass}, 
Phys. Rev. 125 (1962) 397; 128 (1962) 2425.}
\nref\rgel{M. Gell-Mann, {\it A Schematic Model of Baryons and
Mesons}, Phys. Letts. 8 (1964) 214.}
\nref\rwil{K. Wilson, {\it Confinement of Quarks}, 
Phys. Rev. D 10 (1974) 2445.}
\nref\rgalb{Galileo Galilei, {\it The Dialog Concerning the Two World
Systems}, (Univ. Calif. Press 1967), Trans. Stillman Drake}
\nref\rbacc{F. Bacon,{ op. cit.}, p.102}
 
\Title{\vbox{\baselineskip12pt
  \hbox{ITPSB 96-56}
  \hbox{hepth 9609XXX}}}
  {\vbox{\centerline{Modern Metaphysics}}}

  
  \centerline{ Barry~M.~McCoy~\foot{mccoy@max.physics.sunysb.edu}}

  \bigskip\centerline{\it Institute for Theoretical Physics}
  \centerline{\it State University of New York}
  \centerline{\it Stony Brook,  NY 11794-3840}
  \bigskip
  \Date{\hfill 09/96}

  \eject

\centerline{\bf Abstract}

Metaphysics is the science of being and asks the question ``What
really exists?'' The answer to this question has been sought for by
mankind since the beginning of recorded time. In the past 2500 years
there have been many answers to this question and these answers
dominate our view of how physics is done. Examples of questions which were
originally metaphysical are the shape of the earth, the motion of the
earth, the existence of atoms, the relativity of space and time, the
uncertainty principle, the
renormalization of field theory and the
existence of quarks and strings. I will explore our changing
conception of what constitutes reality by examining the views of
Aristotle,  Ptolemy, St. Thomas Aquinas, Copernicus,  Galileo, 
Bacon, Descartes, 
Newton, Leibnitz, Compte, Einstein, Bohr, Feynman, 
Schwinger, Yang, Gell-Mann, Wilson and Witten.

\newsec{ Introduction}

In recent years Physics has come under attack from politicians and from
many in the general public. The funding for the Superconducting Super
Collider was terminated, job prospects are drying up, and nuclear physics
has become so unpopular as a result of the nuclear power debate that
the very word nuclear has been dropped from Nuclear Magnetic Resonance
in order that the general public will accept Magnetic Resonance Imaging
as a medical tool.

This hostility to physics is not a new phenomena and has very deep and
ancient roots. For example at the end of  Roman Empire St. Augustine
wrote~\raug~

{\it Th good Christian should beware of mathematicians, and all those
who make empty prophecies. The danger already exits that the
mathematicians have made a covenant with the devil to darken the
spirit and to confine man in the bonds of Hell.}

Somewhat more recently, in 1511 during the Renaissance, Erasmus wrote in ``In
Praise of Folly''~\reras~

{\it Near these march the scientists, reveranced for their beards and
the fur on their gowns, who teach that they alone are wise while the
rest of mortal men flit about as shadows. How pleasantly they dote,
indeed, while they construct their numberless worlds, and measure the
sun, moon, stars and spheres as with thumb and line. They assign
causes for lightening, winds, eclipses and other inexplicable things,
never hesitating a whit, as it they were privy to the secrets of
nature, artificer of things, or as if they visited us fresh from the
council of the gods. Yet all the while nature is laughing grandly at
them and their conjectures. For to prove that they have good
intelligence of nothing, this is a sufficient argument: they can never
explain why they disagree with each other on every subject. Thus
knowing nothing in general, they profess to know all things in
particular; though they are ignorant even of themselves, and on
occasion do not see the ditch or the stone lying across their
path. because many of them are blear-eyed or absent minded; yet they
proclaim that they perceive ideas, universals, forms without matter,
primary substances, quiddities and ecceities--things so tenuous, I
fear, that Lynceus himself could not see them. When they especially
disdain the vulgar crowd is when they bring out their triangle,
quadrangles, circles and mathematical pictures of the sort, lay one
upon the other, intertwine them into a maze, then deploy some letters
as if in line of battle, and presently do it over in reverse
order--and all to involve the uninitiated in darkness. Their
fraternity does not lack those who predict future events by
consulting the stars, and promise wonders even more magical; and
these lucky scientists find people to believe them.}

I do not agree with the conclusions of St. Augustine, Erasmus, and those
who would contract the support for research in physics. On the contrary
 I think that physics and physics education is of great
importance for the general public. In particular I think that  all 
students  can profit greatly from learning the way
that physicists think about the world.

Nevertheless the criticism of Erasmus is still very mush to the point
even though it is almost 500 years old. Physicists still have the
habit of speaking in a specialized language and of relying heavily on
mathematical symbols and arguments. This drastically limits the
audience to which we can communicate our ideas. 

I would like to think that this state of affairs can be improved upon
and  thus my first goal for    
this paper  is an attempt  to see if I can convey the 
essence and importance of physical
thought down through the ages from Aristotle to Witten in a form which
is accessible to people with no laboratory experience and which does
not use any mathematics. In other words I will attempt to explain
physics from a humanist perspective.

My second goal  for this paper is more specialized. I want
 to give an overview of 
the evolution of theoretical physics in the last 50 years from a
discipline which sought guidance from experiments to one which gives
guidance to mathematics. This change is dramatically illustrated by
comparing 1957 with 1990. In 1957 Lee and 
Yang won the Noble prize in physics by
using theoretical methods to explain a puzzle concerning particle
decays which had been found by high energy physics experiments. In
1990 Witten won the Fields Medal in mathematics by using methods of
physics to solve theoretical problems posed by mathematicians. In
both cases the methods used are those of quantum field theory but the  
ends to which those methods are employed are drastically
different. Moreover these two cases are not isolated instances but are
representative of a change in attitude and direction which, in
my opinion, can be rather precisely dated to have begun in 1964 with the paper
by Gell-Mann which introduces the concept of quarks. It is this
historic change of direction of theoretical physics which I had
originally planned  to
discuss in this paper. 

However in preparing this  paper I  have
became aware that in any attempt to generalize about the direction of
research in physics I tread on very dangerous
ground and that there is no way in which I can expect that most of my
colleagues are going to agree with everything  I want to
say. Consequently, in order not to run into semantic difficulties at
the very beginning I  have chosen instead to
write on what I choose to call metaphysics instead of theoretical
physics. In this way if I offend
anyone it will,  I hope, be limited to members of philosophy departments.

\newsec{ Rules for Learning}

The first lesson from philosophy that needs to be understood is that
terrible  misunderstandings arise from the unavoidable fact that
communication requires language and in order for two people to
communicate they must have the same understanding of the words they
use. Unhappily this means that if I am trying to communicate to you an
idea which I have but which you do not then the communication cannot possibly
succeed because we cannot possibly have the same language in
common. Thus it is philosophically impossible for me to teach you
anything new at all. This  phenomena  that teaching is impossible is
encountered by professors 
every day they teach a class.

However, the converse is indeed possible. That is even though I
cannot teach I can learn. This will mean that I will be interpreting
and giving my meaning to the words of others. You can, if you will,
say that I am making a translation from the language of others into my
own language. Sometimes this is translation in a literal sense as when
the original article is written in French, German, Latin or
Greek. Sometimes it is an interpretation from mathematical to physical
language and often it is the translation of ideas. But in all cases the
learning I have in mind is something which occurs in the mind of the
learner and not in the mind of the teacher.    

In order to effectively learn anything it is most helpful for the
learner to have a method.
And  thus I want to begin by outlining some
rules for learning.

First a rule of my own invention:

1)~~{\bf Always look for what is correct in an author.} 

Never adopt the
attitude that if you can find one thing wrong in someone's writing or
thinking that
everything else they write or say is to be disregarded. If you do this you
will soon discover that you cannot read anything because there is no
book or writer that does not contain something which you will call an error.
Instead you must look for what is true and useful. The result is then
that you must judge a writer by his best and ignore his errors in
assessing his reputation.

 Let me give an  example:

Ptolemy was the greatest astronomer of the ancient world. His book
``The Almagest''~\rptol~ written in 150 AD is a scientific masterpiece. He
proves
the world is round even though the direct experimental proof of
Magellan was completed only in 1522. Ptolemy was able to
describe the midnight sun from theoretical reasons alone even though in
150 AD no one who has left a written record had ever been north of
the arctic circle.  Ptolemy understood the precession of the equinoxes
and had a wonderful method of calculating eclipses and the observed
motion of the planets which gave a very precise fit to the data. The
man must be regarded as a genius.

And yet the typical assessment of Ptolemy is that he was a fool
because he thought that the sun moved and the earth stood still. My  rule
says that Ptolemy must be judged for the many things he got  right and
not for the very few things he got wrong.

The second rule is from Descartes in his famous {\it Discourse on
Method}~\rdesc~ written in 1637.

2)~~{\bf Begin by doubting everything.}

This is very easy to say but very hard to put into practice. In
particular you must never accept someone's belief or idea on the basis
of their authority, status, or position in the field. This is
naturally bound to get you into a lot of trouble. It even got
Descartes into a lot of trouble and he lived much of his life
in exile from his native France because of it.

The remaining rules are those given by  Francis Bacon in his book {\it Novum
Organum}~\rbaca~ written in 1620 and are called by him the four idols:

3)~~{\bf Idols of the tribe:}

These are errors of learning which are  inescapably common to all
mankind. They arise because {\it``It is the case that all our perceptions,
both of our sense and of our minds, are reflections of man, not of the
universe, and that human understanding is like an uneven mirror that
cannot reflect truly the rays from objects, but distorts and corrupts
the nature of things by mingling its own nature with it.''}

For example: 

{\it``The human understanding on account of its own nature readily
supposes a greater order and uniformity in things than it
finds. And though there are many things in nature which are unique and
quite unlike anything else, the understanding devises parallels,
correspondences and relations which are not there $\cdots$

 The human understanding, since it has adopted opinions, either
because they were already accepted and believed, or because it likes
them, draws everything else to support and agree with them. And though
it may meet a greater number and weight of contrary instances, it
will, with great an harmful prejudice, ignore or condemn or exclude
them by introducing some distinction, in order that the authority of
those earlier assumptions may remain intact and unharmed $\cdots$

 The human understanding is most moved by things that strike and
enter the mind together and suddenly... .It then imagines that
everything else behaves in the same way as those few things with which
it has become engaged $\cdots$

 The human understanding ... is infused by desire and emotion, which
give rise to 'wishful science' ''}

And finally {\it ``by far the greatest impediment and aberration of the
human understanding arises from the dullness and inadequacy of the senses.''}

4)~~ {\bf Idols of the cave:}

These are errors in learning which are characteristic of individuals
personal ego and education.

In particular:

 {\it ``Men become attached to particular sciences and contemplations
because they think themselves their authors and inventors, or because
they have done much work on them and have become habituated to them.''}

5)~~{\bf Idols of the forum:}

These are errors in learning which are caused by the unavoidable use
of words and the ``alliance of words and names.''  They are of two
kinds.  {\it ``Either they are names of things that do not exist (for just as
there are things without names because they have never been seen, so
also there are names without things, as a result of fanciful
suppositions); or they are names of objects which to exist but are
muddled and vague.''} 

6)~~{\bf Idols of the theater:}

These are errors of learning which are not innate to knowledge itself
but  {\it ``are imposed and received entirely from the fictitious tales in
theories, and from wrong-headed laws of demonstration.''}

For example:

 {\it ``The school of rational philosophers seizes from experience
a variety of common instances without properly checking them, or
thoroughly examining and weighing them, and leaves the rest to
cogitation and agitation of wit.''}
 
On the other hand there  is a  {\it ``class of philosophers 
who, after toiling with great care
and precision over a few experiments, have presumed to devise and
produce philosophies from them, twisting everything else in
extraordinary ways to fit in with them.''}

I rather suspect that in these quotes from 1620 it is possible to
recognize traits that may be seen in your colleagues and which you may
have had to fight against in your own thinking and research.

I will attempt to  use these 6 rules as precepts for what follows
in this lecture.

\newsec{ Natural Science, Mathematics and Metaphysics:Aristotle to Aquinas}

It is a very common literary device to use when writing a paper to
start by referring to something which is ``recent''. Thus Voltaire
in the beginning of the 18th century writes about the ``recent `` work
of Newton; papers on particle physics in the 60's and 70's would talk
about the ``recent''work of Gell-Mann; and today many papers can be
found which start with the formula ``Recently Witten...'' 

But I plan to discuss metaphysics and the very word ``metaphysics''
has gone out of use and out of fashion with physicists hundreds of
years ago. Indeed its use was ridiculed and it was in effect forced
out of physics by a centuries long campaign of abuse. Therefore I
cannot possibly start my discussion by citing ``recent authors'' and
consequently I will of necessity adopt the  opposite strategy of
beginning my presentation with the first author who considered the
topic instead of the most recent. I thus begin by quoting from the
book {\it Metaphysics} by Aristotle~\rari~ written in the fourth century BC.

{\bf There are three kinds of theoretical philosophy; mathematics,
natural science and metaphysics.}

The meanings the words theoretical, mathematics, natural, and
metaphysics have been discussed by some of the most profound
philosophers of 
ancient and medieval times. I will here follow the unsurpassed treatment
given by St.Thomas Aquinas in {\it Expositio super Lebrum Boethii de
Trinitate}~\raqu~ written during 1255--1259 as an explanation of the
book {\it de Trinitate}
written by the Roman philosopher Boethius  in
the sixth century.

St. Thomas begins by quoting Boethius whom I summarize as follows:

{\it ``Come then, let us enter into each matter, discussing it so it can
be grasped and understood, for it seems well said that educated people
try for such certainty as the matter itself allows.}

{\it For theoretical science divides into three---}

{\it 1)~Natural science (physics) which deals with observed matter which
undergoes change. The ideas studied by physics are not abstracted from
observed matter and embody all the changes which matter is subject to.}

{\it 2)~ Mathematics which conceives bodily forms apart from matter and thus
apart from change though the ideas themselves do exist in matter and
so cannot be separated from matter and change.}

{\it 3)~ Metaphysics which is changeless, abstract and separated from observed
matter.}

{\it In natural science we make use of reason, in mathematics we make use
of discipline and in metaphysics we use the intellect which does not
rely on imagination but rather scrutinizes existence itself from which
all existence exists.''}

First we should understand what  St.Thomas means by theoretical:
 
``{\it Speculative or theoretical, as distinct from operative or 
practical, understanding is characterized by attention to truth for
its own sake, rather than as the means to some other activity... .The
subject matter of practical sciences has to be things we can make or
do... .The subject matter of theoretical sciences, on the other hand,
has to be things not made by us, which we cannot be seeking to know
for activities sake.''} Thus St. Thomas makes a distinction between what
we now call pure physics which he calls theoretical and applied
physics which he calls a practical art. Moreover the branches of
philosophy of ethics, aesthetics and politics are 
also regarded by St.Thomas as
practical and not as theoretical branches of knowledge.

Next we need to understand his conception of physics (or natural
science). The role of physics is to comprehend what can be observed
by sensory perception and what we can make images of in our mind
(imagination).  Physics studies observed matter which has
bulk, quantity and can be observed and measured and this matter is in
general not static but is undergoing change. We study matter by using
the processes of reason.
{\it ``Natural science starts from what is more knowable to
us and less knowable in its own nature, using proofs from symptoms and
effects.''}  We derive knowledge of one thing from knowledge of
something external to it---knowledge of effects, for example, from
knowledge of their cause.

This definition  agrees very well with
that of the present day.

Moreover St.Thomas's conception of mathematics is exactly that of the
present day. From observed matter Thomas abstracts the notion of
quantity. {\it ``Quantity, therefore does not depend for its definition on
material--as --perceptible but only on material--as-- thinkable; namely,
substance without its material properties, which is something only
thought can comprehend, and to which our senses cannot
penetrate. Mathematics is the science of   objects abstracted in this
way, and considers only quantity in things and whatever accompanies
quantity; shapes and the like. $\cdots$ In the mathematical science we
argue from definitions of things, proving conclusions by appeal to
formal principles, never deriving truths about something by appeal to
something external to it but by appeal to its own definition.''}
Quoting Ptolemy he says {\it ``Only mathematics, if you examine matters
closely, builds up in its students sure and stable beliefs by means of
irrefutable proofs.''} This is what he means by the discipline which he
says is characteristic of mathematics.

It remains to discuss what St. Thomas means by metaphysics.
 
There are two concepts which are crucial in the understanding of the
metaphysics of Aquinas: 1) incommensurable length scales and 2)
immaterial substances.

By  incommensurable length scales Aquinas means that the ratio of the
length scale of metaphysical objects and phenomena  to the length
scale of observed physical material phenomena is strictly infinite (or
zero depending on how you look at it). 

By  immaterial substance Aquinas means that the fundamental objects on
this infinite metaphysical length scale are not in any direct manner
connected with observed matter on the physical length scale. Indeed,
Aquinas stresses that it is quite inappropriate to even attempt to use
the words and properties which we use to describe observed matter to
describe an immaterial metaphysical substance.

The metaphysics of St.Thomas is the statement that the immaterial
objects of the infinite length scale constitute the unchanging basis
of physical reality and that all  observed material phenomena are to
be derived from them. The process of derivation is through mathematics
much in the sense that mathematics allows us to discuss the concept of
a limit and here again Aquinas is very clear that the properties of
a limit do not have to be the same as the objects through which the limit
is taken.

St.Thomas thus gives a very concrete explanation to his students of
what he interprets Boethius to mean by saying that metaphysics is
changeless, abstract and separated from matter.

This metaphysics  is remarkably sophisticated, and for this
remarkable sophistication  Aquinas was greatly rewarded, He was granted
tenure in this world and when he passed on to the next world the
Catholic church made him a saint, an honor no scientist since his time
has attained. 

But the concepts of infinite ratios of length scales and immaterial substances
which are not the same as observed matter were very hard to swallow.
Moreover, this metaphysics is only what we
would call today a kinematics and does not contain any dynamical
principle. Consequently, although it provides a quite plausible
framework in which to discuss reality it does not provide any tools to
allow the computations of any actual properties or
effects, and in time this metaphysics was slowly abandoned by the
scientific community. Indeed it was more than abandoned it was
ridiculed and castigated and Aquinas himself was subjected to a
remarkable amount of {\it ad hominum} abuse and the development of
scientific thought went in an entirely new direction.

\newsec{ Galileo's revolution of 1610}

The next great advance in science came in
the mid 16th century when Copernicus boldly argued on theoretical
grounds that the earth was
not the center of the universe but that instead the earth moved around
the sun. But even this, epoch making as it was, takes second place to
the totally revolutionary publication by Galileo on March 12, 1610 of
the paper {\it Siderius Nuncius}~\rgala.

It is absolutely impossible to overstate the importance of this paper
of Galileo. In it he announces three things which were totally
unanticipated and epoch making:

{\bf 1.~The invention of the telescope;}

{\bf 2.~The observation of mountains on the moon;}

{\bf 3.~The discovery of four moons of Jupiter;}

Even after the passage of 386 years the excitement of Galileo is
infectious:

{\it ``In this short treatise I propose great things for inspection and
contemplation by every explorer of Nature. Great, I say, because of
the excellence of the things themselves, because of their newness,
unheard of throughout the ages, and also because of the instrument
with the benefit of which they make themselves manifest to our
sight.}

{\it Certainly it is a great thing to add to the countless multitude
of fixed stars visible hitherto by natural means and expose to our
eyes innumerable others never seen before, which exceed tenfold the
number of old and known ones.}

{\it It is most beautiful and pleasing to the eye to look upon the
lunar body, distant from us about sixty terrestrial diameters, from so
near as if it were distant by only two of those measures, so that the
diameter of the same moon appears as if it were thirty
times... larger than when observes only with the naked eye. Anyone
will then understand with the  certainty of the senses that the moon
is by no means endowed with a smooth and polished surface, but is
rough and uneven and, just as the face of the Earth itself, crowded
everywhere with vast promontories, deep chasms, and convolutions.}

{\it But what greatly exceeds all admiration, and what especially
impelled us to give notice to all astronomers and philosophers, is
this, that we have discovered four wandering stars, known or observed
by no one before us... All these things were discovered and observed
a few days ago by means of a glass contrived by me after I had been
inspired by divine grace.''}

Galileo's revolution has been so complete that the statement I 
gave of it  does not even sound revolutionary. What he did was to
built an experimental apparatus and use it to make an observation
which had not been possible before that apparatus was built. We do not
think today that this is strange. What is revolutionary in Galileo's
work is the  {\bf this is the first time observations with instruments
had ever been done.}

The metaphysical consequences of this paper were profound and
immediate. For all of previous history the words {\it sensory
perception}
had meant {\bf unaided} sensory perception. The immediate
metaphysical question to answer was this:
{\bf Are observations made with instruments to be considered as being
real?} In other words, were the moons of Jupiter really out there in
the heavens or were they inside the telescope of  Galileo. However,
unlike all previous metaphysical questions this one was answered
within a year. The universal answer was that if the observations with
instruments could be repeated by others then the phenomena had just as
much status to the title of reality as any observation which did not
involve instruments.

With the invention of the telescope science underwent a permanent
change. Suddenly improvements in technology meant that logic and 
deductive reasoning were not the only way to learn about
nature. Anyone who could build a better lens or microscope could go
and point it at something and make a new discovery without paying any
attention to the theory at all. And thus the metaphysics and definition
of reality of Aquinas was abandoned in a mad rush. With the invention
of the telescope  the dominance of theoretical over
experimental methods which had existed for almost 2000 years was overthrown.
Why should anyone worry about the true definition of reality
and being when there were new planets and new biology to discover?

\newsec{ The experimental metaphysics of Bacon} 

Such a profound revolution in technology demanded an equally profound
revolution in the metaphysical basis of science. This was provided
within 10 years by Bacon in the same book which contained the four
idols we talked about before. In {\it Novum Organum} Bacon sets forth a
scientific method which is diametrically opposed to the metaphysical
conception of reality of Aquinas. Bacon's method has been so
universally accepted that for generations it has been called {\it the}
scientific method. It is taught in our elementary and secondary
schools as absolute truth. It has caused the very definition of
physics to go from a theoretical science to an experimental science.

I quote again from {\it Novum Organum}~\rbacb

{\it ``Now the directions for the interpretation of Nature are of two
separate kinds: the first for eliciting or devising axioms from
experience, the second for drawing or deriving new experiments from
axioms. the former again is divided three ways, that is into three
provisions: that for the sense, for the memory, and for the mind or
reason.}

{\it First of all, a sufficient and suitable natural and experimental
history must be compiled. That is fundamental to the matter. For
there must be no imagining or supposing, but simply discovering, what
nature does or undergoes.}

{\it But this natural and experimental history is so various and
scattered that it would confuse and distract the understanding, unless
it is set out and presented in a suitable order, for which purpose
table and arrangements of instances should be drawn up, and put
together in such a manner and order as to enable the understanding to
deal with them.''} Only then are we able to {\it ``employ a legitimate and
true induction, which is the very key of interpretation.''}

Bacon~\rbacd~ makes very explicit the relation which he thinks his method has
to the incommensurable length scales of Aquinas:

{\it ``In this way we shall be led, not to the atom, which presupposes a
vacuum and immutable substance (both of which are false) but to real
particles such as are found. Nor again is there cause for alarm at the
subtlety of the inquiry, as if it were inexplicable; on the contrary
the closer the inquiry comes to simple natures, the more intelligible
and clear will everything become; the business will be transferred
from the complicated to the simple, from the incommensurable to the
commensurable, from the irrational to the rational, from the
indefinite and doubtful to the definite and certain.''} 

 There is no place in Bacon's theory for infinite
length scales and certainly there is no place for ``immaterial
substances.''
For Bacon the question of reality and being is self evident. If you
can perceive it and measure it it is real. Otherwise don't talk
about it. Or as Bacon puts it~\rbace~ {\it ``there are two practical division in
science; physics corresponds to the mechanical arts; metaphysics
corresponds to magic.''}

Bacon resoundingly places experiment above theory.
 
Bacon was also extremely successful with his method. In his book 
Bacon applies his method to the question of heat  and
concludes by finding that {\bf heat is motion.} In other words, in
1620 Bacon invented the kinetic theory of gases and heat on the basis
of analyzing the experimental evidence of his day. Truly an
achievement of genius.

\newsec{The gravitation, vacuum and particles of Newton}

But an even greater creation of genius is the invention of the
universal theory of gravitation by Newton as published in the {\it
Principia} of 1687. I think it is fair to say that no scientist in
history has so deeply impressed the general public. As proof of this I
offer a description of Newton's discovery written not by a scientist
but by one of the greatest of all writers of the 18th century, Voltaire~\rvol

{\it ``He (Newton) said to himself: 'From whatever height in our
hemisphere these bodies might fall, their fall would certainly be in
the progression discovered by Galileo, and the spaces traversed by
them would be equal to the squares of the time taken. This force which
makes heavy bodies descend is the same, with no appreciable
diminution, at whatever depth one may be in the earth and on the
highest mountain. Why shouldn't this force stretch right up to the
moon? And if it is true that it stretches as far as that, is it not
highly probable that this force keeps the moon in its orbit and
determines its movement? But if the moon obeys this principle,
whatever it may be, is it not also very reasonable to think that the
other planets are similarly influenced.}

{\it If this force exists it must increase in inverse ratio to the
squares of the distances. So it only remains to examine the distance
covered by a heavy body falling to the ground from a medium height,
and that covered in the same time by a body falling from the orbit of
the moon. To know this it only remains to have the measurements of the
earth and the distance from the earth to the moon.'}

{\it This is how Newton reasoned. But in England at that time there
existed only very erroneous measurements of our globe...As these false
calculations did not agree with the conclusions Newton wanted to draw,
he abandoned them. A mediocre scientist, motivated solely by vanity,
would have made the measurements of the earth fit in with his system
as best he could. Newton preferred to abandon his project for the time
being. But since M. Picard had measured the earth accurately by
tracing this meridian, which is such an honor for France, Newton took
up his first ideas again and found what he wanted in the calculations
of M. Picard. This is a thing that still seems admirable to me; to
have discovered such a sublime truths with a quadrant and a bit of
arithmetic.''}

But, unlike the kinetic theory of heat, whose
explanation has not changed since the time of Bacon,  this gravitational
attraction which is an instantaneous action at a distance was
and is metaphysically unsettling. And Newton knew it. To quote
Voltaire again:
 
{\it ``Newton foresaw clearly when he had demonstrated the existence of
this principle that people would revolt against its very name. In more
that one place in his book he cautions the reader against gravitation
itself and warns him not to confuse it with what the ancients termed
occult qualities, but to be satisfied with the knowledge that there is
in all bodies a central force which acts from end to end of the
universe   on the nearest and most distant bodies in accordance with
the changeless laws of mechanics.''}

Indeed, Newton was right. In theoretical and 
metaphysical terms this action at a
distance was very hard for some to swallow. One of those who would not
swallow it was Leibnitz who for his entire career argued for what we
now call a principle of locality. Leibnitz~\rleib~ in his paper which is in
reply to the {\it fourth paper } of Clarke says
{\it ``It is a strange fiction to regard all matter as having gravity,
and even to regard it as gravitating towards all other matter, as if
every body had an equal attraction for every other body in proportion
to mass and distance; and this by means of attraction properly so
called, and not derived from an occult impulsion of the
bodies. Whereas in truth the gravitation of sensible bodies towards the
center of the earth must be produced by the movement of some fluid. And
the same is true of other gravitations such as those of the planets
towards the sun or towards one another. A body is never moved
naturally except by another body which impels it by touching it; and
after this it goes on until it is hindered by another body touching
it. Any other operation on bodies is either miraculous or imaginary.''}

 Leibnitz thus is in explicit contradiction with Newton and condemns
the hypothesis that the gravitational attraction of bodies could proceed
instantaneously without a time delay. Instead he believed that forces
could only act by direct contact. It is difficult to call him wrong but
he had no competing theory that would make predictions of planetary 
motions. Consequently Leibnitz
ran into the same sort of abuse which Aquinas did and at the hands of
Voltaire suffered tremendous personal attacks  and condemnation. His
reputation survived because he was, after all, one of the inventors of
the calculus and because he could compute as well as Newton. But the
credit in the 18th century for the revolution in human thought  caused
by the invention of the science of mechanics  was all given to Newton.

If action at a distance was the only metaphysical invention of
Newton that would already be profound but in fact Newton had a much
larger impact than even gravity would indicate and another of the key results
which the 18th  century credited Newton was the demonstration of the
existence of the vacuum.

At the time of Newton it was hotly debated whether space was filled
with continuous matter or whether there was such a thing as empty
space. Descartes, for example, argued that the universe was filled with some
substance, called a {\it plenum} and that objects such as the planets
consist of vortices in this substance.  In contradiction to this was
the metaphysical idea of Newton that there was such a thing as
empty space and that reality consisted of very tiny
material particles which moved in this space. The metaphysics of
Newton is diametrically opposed to Descartes (and indeed also is
opposed to Bacon). The history of 18th
century physics is the triumph of Newton's ideas of gravity, empty
space, atomism at the expense of continuous matter and local interactions.   

\newsec{Fields and Relativity}

However the metaphysics of locality and continuity had an appeal that
did not die even with the enormous success of Newtonian mechanics and
indeed proved to be exactly the metaphysics needed by the 19th century
for the study of electromagnetism. It had been known from the
beginning of the 18th century with the work of Roehmer that the
velocity of light was finite and that light did not have an infinite
velocity like the action at a distance of Newtonian gravity. The
theoretical work of Faraday and Maxwell created for electromagnetism a
metaphysical framework that threw out everything which Newton had
used. Faraday and Maxwell described electromagnetism in terms of a
field which is present at every place in space. It is this field which carries
radio and television signals.   This field is viewed
as real and interactions with the field are local in exactly the sense
that Leibnitz meant by the word local. 

The completion of the destruction of Newtonian metaphysics was carried
out by Einstein. In the general theory of relativity
Einstein not only replaced the vacuum by fields but
replaces the gravitational action at a distance in the static unchanging
space of Newton by a dynamic space where geometry itself changes. 

But neither Faraday, nor Maxwell, nor Einstein found it necessary to 
eliminate the notion of a particle. Thus at the beginning of the 20th
century both particles and fields were considered to be real and were
considered to be appropriate for the description of very different
things. 

\newsec{The observability of Compte and Bohr}

From Galileo, Bacon and Newton the metaphysics of the experimental
definition of reality continued to grow literally without bound until
in the beginning of the 19th century Compte codified it into the
philosophy of {\bf positivism} in which the only things that are entitled to
the status of {\it reality} are those which can be positively
measured. One description of this philosophy is~\rrog:

{\it ``Positivism means the definite abandonment of all search for
ultimate causes, and the turning of human attention rather to the laws
of phenomena as the only kind of knowledge which is both attainable and
useful. Knowledge is of value to us because it helps us modify the
conditions in the physical and social world; to do this we need to
know how things act, and that is all we need to know.''}

Metaphysics has been so far degraded by the time of Compte
that the term ``metaphysical'' is used as a pejorative to describe
one of the former eras of primitive thought from which we have now
happily emerged into the light of pure reason. Science to Compte is
the correlation of phenomenological observations between complex
systems. His greatest achievement was to apply this vision of science
to the most complex system he could think of-- Society itself-- and
is credited with being the father of sociology. Soon thereafter Marx
took the identical conception of science as the abstracting of laws
made from the observation of complex systems and applied it to
economics which lead to the creation of what has since been called
``Scientific Marxism.''

The ultimate expression of this evolution in the definition of reality
by means of experimental observation is the invention and
interpretation of quantum
mechanics at the hands of Bohr and the Copenhagen school. Thus 
by the 1920's reality and being are defined in terms of {\bf
observable.} The very word defines the metaphysics to be used. The
triumph of quantum mechanics in explaining atomic spectra and the
success of the uncertainly principle were seen by the 1930's to be a
complete vindication of the doctrine that only the observed can be
called real. Indeed we teach this in our quantum mechanics courses
every day of the week.

\newsec{The revolution of 1964}

By the end of the 1930's it is fair to say that absolutely no
physicist wanted to challenge this observationally and experimentally
based conception of reality. After all it had explained so much which
had been thought to be incomprehensible before. Yet there was one
nagging problem. When you applied the rules to some questions you
sometimes got infinite answers. 

The resolution of this problem with infinities was given in 1948 by
Feynman, Schwinger and Tomonaga. We can now interpret their procedure
as a breakdown of the metaphysical premise that only the observed is
real but it was certainly not viewed as such at the time. It was
merely a useful device for getting rid of a nagging problem. But
because it is in fact the precursor to a much bigger shift in point of
view it is most helpful to present this invention of renormalized
quantum field theory from a metaphysical rather than from a
computational perspective.
  
Reality in 1948 consisted in several metaphysical assumptions: 

1) Space is continuous and, even more strongly, all physically observed
properties had to obey the Einstein laws of special relativity;

2) Particles are observable in accelerator experiments and are in a
close (usually one to one) correspondence with fields. Fields and particles
were considered to be interchangeable from the point of view of ``reality''.

3) From these fields there was constructed a dynamical principle 
which allowed mathematical calculations to be done. These interactions
were considered to be local in that the fields depended on a single
space-time point and only products of fields at the same space-time
point were allowed.

 It was 
calculations  coming from these three principles 
that sometimes gave infinite results.

In effect, though this is not how it was viewed at the time, Schwinger,
Feynman and Tomanaga, modified the metaphysics of this situation by
replacing the first metaphysical principle and the easiest explanation
of their replacement (though this is not technically what they did) is
to replace continuous space with a lattice with a spacing $a.$ Then
all computations were carried out on this fictitious conception of
space. However, it was realized that the scale of laboratory
observations could not possibly be the scale of this fictitious
lattice and at the end of the calculation the theory was rescaled to
go from the scale of the lattice to the scale of the laboratory and
then a limit was taken that made the lattice spacing go to zero in
relation to the laboratory length scale which was kept finite. This
process of {\bf changing the normalization of the length scale is
called renormalization}

This procedure gave finite answers where before there were
infinities and is one of the greatest triumphs of 20th century physics.  

{\bf This procedure will also be recognized as precisely the idea of
the incommensurability of the metaphysical and the observed length
scale introduced by Aquinas in 1255.}

Of course this interpretation of the theory renormalization of quantum
electrodynamics as the reinvention of one of the key parts of the
metaphysics of Aquinas was not made in 1948. 
And moreover the invention of renormalization had 
only destroyed one of the metaphysical
principles of quantum field theory of 1948. But even renormalization
theory retained the more or less direct connection between particles and
fields. As very important examples I need only point out the famous
 1954 paper of Yang and Mills~\rym~ on non Abelian gauge theory
where the proton and neutron are each represented by their own field
and the equally famous paper of Lee and Yang~\rly~ of 1956 where parity
violation in weak interactions are explained by a theory which again
has a one to one correspondence between fields and particles. 

But this one--to--one correspondence of particles with fields was not
to last for long and the first suggestion that it contains
contradictions and paradoxes was made by Schwinger~\rschwa~in the
preface to his book {\it Selected Papers in Quantum Electrodynamics;}

{\it ``Thus, although the starting point of the theory is the
independent assignment of properties to the 
[electromagnetic and electron] fields, they can never be disengaged
to give those properties observational significance. It seems that we
have reached the limits of the quantum theory of measurement, which
asserts the possibility of instantaneous observations, without
reference to specific agencies. The localization of charge with
indefinite precision requires for its realization a coupling with the
electromagnetic field that can attain arbitrarily large
magnitudes. The resulting appearance of divergences, and
contradiction, serves to deny the basic measurement hypothesis. We
conclude that a convergent theory cannot be formulated consistently
within the framework of present space--time concepts.''} This is a
direct and explicitly stated challenge to the Baconian  metaphysics
of observability. Furthermore Schwinger went on to illustrate his
philosophical idea by inventing a model in one space and one time
dimension ~\rschwb~where an
electromagnetic field and a massless electron field do indeed merge to 
yield one single massive excitation.

But the full attack on the one--to--one correspondence in made in 1964 by
Murray Gell--Mann~\rgel~ in the paper
``A schematic model of baryons and mesons'' where 
the spectrum of strongly interacting particles is explained  
by introducing the notion
of a quark.

I have now arrived at the dangerous part of my talk. As long as I have
talked about people who are either dead or whose papers have clear and 
incontestable correct statements I am on fairly safe ground. But with
the paper of Gell-Mann I am dealing with someone very much alive. As
to clarity I will quote the author himself and let the reader judge.

Gell-Mann wanted to explain why the strongly interacting baryons and
mesons have masses which are related by a group theoretic
symmetry. For our consideration the mathematics of the symmetry is not
important. What is important is that the observed particles were not
the simplest way to realize the symmetry and that the simplest
explanation of the masses of the observed strongly interacting
particles was obtained in terms of three hypothetical objects which he
called ``quarks'' after a reference he quotes from {\it Finnegans Wake}
by James Joyce. These hypothetical objects have fractional charges of
${-{1\over 3}e}$ and ${+{2\over 3}e}.$ He then writes the following 2
sentences in the last paragraph:

{\it ``It is fun to speculate about the way quarks would behave if they
were physical particles of finite mass (instead of purely mathematical
entities as they would be in the limit  of infinite mass)''}
  
{\it ``A search for stable quarks of charge ${-{1\over 3}}$ or
${+{2\over 3}}$ and/or stable di-quarks of charge ${-{2\over 3}}$ or
${+{1\over3}}$ or ${+{4\over 3}}$ at the highest energy accelerators
would help to reassure us of the non-existence of real quarks.''}

If Aquinas felt that he had to devote an entire paper to an
explanation of a few lines of Boethius it is not out of place to try
to find out what Gell-Mann was talking about.

To begin with it is clear that he uses the most profound and
debated words of metaphysics; {\bf non-existence} and {\bf real} and
makes the  extremely metaphysical distinction of {\bf physical
particles} as opposed to {\bf hypothetical mathematical objects.} It
is equally clear that he never defines what he means by these words.

One not implausible inference is that Gell-Mann was on the horns of a
dilemma.

 The dilemma is as follows. The theory as proposed by Gell-Mann is based
on fields called quarks, but at the time of the proposal in 1964 no
particle had ever been found that corresponded to the field. Gell-Mann
on the one hand wanted to follow Baconian metaphysics and assert that
quarks were real if they could be observed in high energy accelerator
experiments. On the other hand he clearly seems to be worried that they
would not be detected in these experiments.

In fact particles which would correspond to these quark fields have
never been found. But nevertheless Gell-Mann's suggestion of quark
fields was not rejected. Instead a truly amazing thing
happened. Physicists abandoned the second of the 1948 metaphysical
principal  and accepted the idea that
 
{\bf Fields and particles, not only are not in one to one
correspondence but are completely different concepts.}

In particular:

1)  Fields are considered to be fundamental and the observed particles
are considered to be complex excitations of these fields;

2) Protons, neutrons, pi mesons and other strongly interacting
particles are indeed  observed as particles but there are
no such thing as proton, neutron or pi meson  fields; 

3) Quark fields are  considered as real even though there is no
such thing as an observed quark particle. This phenomena of a field
with no corresponding particle is referred
to as {\bf confinement.}

But  this concept of a field with no material particle to go
with it is exactly the concept of the {\bf immaterial substance as
discussed by Aquinas in 1255.}

This point of view was so rapidly adopted that by 1974 in the paper 
``Confinement of quarks''  Ken Wilson~\rwil~ writes
{\it ``The success of the quark-constituent picture both for resonances
and for deep-inelastic electron and neutrino processes makes it
difficult to believe quarks do not exist. The problem is that quarks
have not been seen. This suggests that quarks, for some reason, cannot
appear as  separate particles in a final state.''} He then goes on to
credit Schwinger~\rschwb~ with exhibiting a confinement mechanism in
1962
and then proposed his own mechanism
based upon a lattice realization of space on the metaphysical scale.
Thus 
 from 1948 to no later that 1974 physics abandoned the
metaphysics of that reality is observation which had ruled scientific
thought since 1610.

\newsec{The rise of the new metaphysics}

The new metaphysical age which physicists have entered since 1964 is
immeasurably richer and  productive than the metaphysics abandoned in
1610. We have indeed returned to the kinematic notions which Aquinas
had of the incommensurate length scales and the immaterial substance
but we now also have quantum dynamical principles from which calculations
can be made. 

But even more than that has occurred. The very  balance between 
experimental and
theoretical thought has been profoundly altered in the past 30 years.
It is no longer true that theorists wait hungrily for data and pounce
on every bump in a cross section measurement. And while it remains as
true as ever that new experiments can give as great a  revelation as
the discovery of the moons of Jupiter it is no longer the case that
the discovery of a new comet or asteroid sets off a wave of theoretical
computations. 

The new metaphysics has made it respectable to once again ask a purely
theoretical question. Or to put it in another fashion, theorists now
cannot escape the fact  that over the decades and centuries
there have piled up an enormous number of observed phenomena which we
do not have a really good theoretical explanation for. Once the
metaphysics which is dominated by experiment is abandoned it is no
longer acceptable for theorists to hide behind the claim of insufficient
experimental data when they have failed to construct an adequate
theory. It is no longer acceptable to conclude a theoretical paper by
pushing the mass of some new presumed particle up higher than the
current limit of observation and then calling upon people to build a
bigger accelerator to test the conjecture. Theoreticians are being
forced to answer theoretical questions in theoretical terms.

Let me give an example of a purely theoretical question for which
there can be no experimental answer. We all know that quantum
mechanics is invented to make atoms stable and that Fermi statistics
are needed to make bulk matter exist. This was mathematically proven
in the late 60's. But it is a slight scandal that we have no proof yet
that organic chemistry exists. By this I mean the following. The
physics which governs organic chemistry is believed to be the 
non relativistic quantum mechanics of atomic nuclei (treated as points)
and electrons interacting with the coulomb interaction. This interaction
is invariant under parity. But all large organic molecules like DNA
are twisted and maximally violate parity conservation. How does this
happen?

To be more precise we note that a key feature of experimental reality
it that the mass ratio of the proton to the electron is 1836:1. It is
plausible that this large mass ratio can be used to make valid the
ball and stick models of organic molecules which are commonly used to
illustrate the twisting of DNA. But surely if the mass ratio is 1:1
then organic chemistry as we know it is impossible. So my purely
theoretical question is this; how large does the  mass ratio have to
be before organic chemistry happens? This is an explicit theoretical
question which cannot be answered by experiment.

Moreover, once you abandon the pre 1948 metaphysics 
and accept infinite length
scales and immaterial substances there is no reason to keep any of the
old metaphysics at all. In particular there is absolutely no longer
any reason to limit yourself to the belief that fields interact only
at points because after all these fields are in the metaphysical space
and as Aquinas emphasized our notions taken from the world of
observation need not be valid on the length scale of the fundamental
metaphysical objects. Therefore in the last 25 years we have invented 
string theory which abandons the principle of
point like interactions of fields  and replaces the concept of the
fundamental point with that of the fundamental string. In other words
the fundamental objects of reality are extended instead of point like. 

Indeed there is not even any reason to suppose that the metaphysical
space has the same number of dimensions as observed space and it is
now perfectly acceptable to talk of metaphysical spaces of 10, 11, or
26 dimensions just so long as they all give an 
observed spacetime of 4 dimensions. 

In order to put string theory in perspective I turn 
once again to a book of Galileo. This
time his famous book {\it The Dialogue Concerning the Two World
Systems}~\rgalb~ published in 1632. This book presented the difference
between the view of Ptolemy, that the earth did not move, and the view
of Copernicus that the earth moved around the sun. At the time both of
these views could easily be called different metaphysics. Both lead to
predictions and explanations of the observed data. 
 
Galileo set out to apply this new point of view of Copernicus to as much 
observational reality as he could
find. But he found that a challenge to an accepted  scientific point
of view is rough going. He was charged with heresy, hauled before the
Roman inquisition, tried and convicted of having the wrong views. This
trial
is justly famous and we now recognize that even though condemned he
was correct. But what is often forgotten is that at the time in 1632
{\bf Ptolemy's theory fit the data better that Copernicus' did. }
Moreover, some of Galileo's arguments such as the statement that the
moon had nothing to do with the tides are just flatly wrong. In 1632
it was not clear which world view was correct.

The same can be said about string theory today. The conventional
standard model of quantum field theory based on the immaterial objects
of quarks and gluons and using the dynamical principle of the Yang
-Mills interaction
 fits all the observed data.
Moreover it is quite possible that there are many 
mistakes in our understanding
of how to get observable predictions from the metaphysical strings.
Just as Galileo needed to abandon his assumption of circular orbits
before he could get better agreement with the data that  the Ptolemaic
theory would give, so string theorists surely need better computational tools
to make predictions that can give decisive tests between the two world
views. At present it is not established which world order on the
metaphysical level better describes observable reality. 

The relation between  experimental physics, mathematics and
metaphysics is now very usefully described by the classification given
by Aquinas in 1255. Experimental physics studies observed properties
and the relations between them. Metaphysics posits the world of
ultimate unchanging reality and mathematics is the connection between
the two. If the application of mathematics to the metaphysical does not
yield observation then we must change the metaphysics. We may never
attain complete knowledge of the metaphysical world but we can and do
improve our comprehension of it.

The new metaphysics of Aquinas thus puts a much greater 
emphasis on mathematics that did the metaphysics of Bacon. In the last
30 years the needs of making the connection between the metaphysical
and the observed has required the invention of much new mathematics. 
This new mathematics turns out to
be deeply related to both statistical mechanics on the one hand and to
topological and algebraic problems in mathematics on the other. It is the
mathematical progress that came from this new physical viewpoint for
which Witten was awarded the Fields Medal in 1990. These advances
would not have been possible had experimental data been allowed to
continue to be the sole arbiter of our theoretical thought.

\newsec{ Conclusion}

I have now concluded my journey through the history of physics,
mathematics and metaphysics. I hope I have made it clear that the
easiest and clearest way to understand both the history and present
day developments in the theoretical sciences is not to focus on the
details of observational experiments or to focus on the details of
mathematical computation but to focus on the metaphysical conception of
reality which is being used. Because whether or not we are willing to
admit it in so many words all scientific research is completely
determined by the metaphysical principle of the person doing the
research. Each one of us has his/her own metaphysical conception of
reality which guides our actions. Each of us profoundly believes in
his/her own metaphysics. But no two of us have exactly the same
metaphysics in common.

Indeed while the details of experiments and mathematical computations
will forever be the province of those few who choose to specialize in
them the metaphysics of science is comprehensible  to everyone. Indeed
metaphysics is not the property of scientists alone but in truth each
and every person under the sun has their own unique set of
metaphysical beliefs. These are the questions which people have
thought about from the beginning of time. Metaphysics, in its own
individual way, is a universal language. 

But if it is true that  physics has dealt with 
questions of great popular interest for 2500 years then why 
do we lament for the lack of jobs and
lack of federal support? Why do so many people seem hostile to science?

This question is also not a new one. And thus I will close 
by again quoting Francis Bacon~\rbacc~ from 1620:

{\it  Moreover, even if such hostility were to cease, the growth of
the sciences would still be hindered by the fact that effort and hard
work in that direction go unrewarded. For those cultivating the
sciences and those paying for them are not the same people. For
scientific advances come from great minds, whereas prizes and rewards
of science for this knowledge are in the hands of the common people, or
leading citizens, who only occasionally are even moderately
educated. Advances of this kind not only go unrewarded with prizes and
substantial benefits, but do not even enjoy popular esteem. For they
are beyond the grasp of most people, and are easily overwhelmed and
extinguished  by the winds of common opinion. It is no wonder if such
an enterprise which is not honored does not prosper.}

And he should know about politics better that almost any scientist in
history because he was not only a great philosopher but was also
 the Lord High Chancellor of England.

\bigskip
{\bf Acknowledgments}

This work was partially supported by NSF grant DMR 9404747.

\vfill
\eject
\listrefs

\vfill\eject

\bye
\end